\documentclass[twocolumn]{article}
\usepackage[pdftex]{graphicx}
\usepackage{url}
\usepackage{subcaption}

\begin{document}
\noindent This paper is the preprint version of the paper ``Catching Unusual Traffic Behavior using TF--IDF-based Port Access Statistics Analysis'' published by IEEE with DOI: 10.1109/CCCI52664.2021.9583212.\\

\noindent\copyright{} 2021 IEEE. Personal use of this material is permitted. Permission from IEEE must be obtained for all other uses, in any current or future media, including reprinting/republishing this material for advertising or promotional purposes, creating new collective works, for resale or redistribution to servers or lists, or reuse of any copyrighted component of this work in other works.\\
\newpage
\title{Catching Unusual Traffic Behavior\\
    using TF--IDF-based Port Access Statistics Analysis}
\author{Keiichi Shima\\
Ineternet Initiative Japan}
\date{\empty}

\maketitle

\begin{abstract}
  Detecting the anomalous behavior of traffic is one of the important actions for network operators. In this study, we applied term frequency -- inverse document frequency (TF--IDF), which is a popular method used in natural language processing, to detect unusual behavior from network access logs. We mapped the term and document concept to the port number and daily access history, respectively, and calculated the TF--IDF. With this approach, we could obtain ports frequently observed in fewer days compared to other port access activities. Such access behaviors are not always malicious activities; however, such information is a good indicator for starting a deeper analysis of traffic behavior. Using a real-life dataset, we could detect two bot-oriented accesses and one unique UDP traffic.
\end{abstract}

\section{Introduction}
Traffic monitoring and anomaly detection are vital operations for network operators. If we see access to a certain port that has not been seen before, then we may infer it to be a new type of attacking or an attacking symptom. Of course, judging whether such activity is malicious behavior is not always possible. It may possibly be simply a random side effect caused by normal operations or mistakes. However, despite the fact that whether that activity is a real attack, it is important for network operators to notice such a behavioral change as soon as possible.

There have been many past approaches that inform us of such a change. The simplest mechanism is the time-series traffic volume-based anomaly detection. It works well when we want to detect a large-scale traffic anomaly such as DoS/DDoS; however, detecting smaller malicious activities is difficult. We have recently seen many different types of attacks to/from IoT devices, which may not always generate considerable traffic but tries reaching a wide range of networks for a long time. Such activity is sometimes buried under the other larger traffic and is difficult to observe. It may be possible to monitor traffic per port and watch the time-series behavior for every port; however, the number of ports is 65,536 in theory and the cost of tracing all the ports is not trivial.

In this paper, we seek a method to obtain irregular behavior of port access statistics using a natural language processing scheme, term frequency -- inverse document frequency (TF--IDF). With this approach, we could distinguish interesting port access activities from other activities.

\section{TF--IDF for Port Access Statistics}

TF--IDF is a popular method for obtaining important words from documents based on a statistical analysis of the occurrence of each word.  Equation~\ref{eqn:tf-idf} shows the formula for calculating a TF--IDF value of a word $t$ in document $d$.

\begin{eqnarray}
    \label{eqn:tf-idf}
    tf(t,d) &=& \frac{n_{t,d}}{\sum_{i\in{}d}n_{i,d}} \nonumber \\
    idf(t) &=& \log\frac{N}{df(t)+1} + 1\nonumber \\
    tfidf(t,d) &=& tf(t,d)\cdot{}idf(t)
\end{eqnarray}

where $t$ denotes a word, $d$ is a document, $n_{t,d}$ is the number of occurrences of the word $t$ in document $d$, $N$ is the total number of documents, and $df(t)$ is the number of documents containing the word $t$. $tf(t,d)$ indicates the popularity of the word $t$ in the document $d$. $idf(t)$ indicates how rarely the word $t$ is present among all documents.

We reuse this idea for the analysis of port access statistics. A port number is considered as a ``word'', and the number of access to the port is considered to be the number of occurrences of the word. The port access statistics for one day are treated as one ``document''. Thus, a document can contain a maximum of 65,535 different words. The number of documents is the same as the number of days we collected the port access statistics. If we obtain considerable access to a specific port, the $tf$ value then increases, indicating that the port is more popular than other ports. However, if the port is observed on many different days, the $idf$ value of the port becomes smaller, and the final $tfidf$ value will be small.

\section{Dataset}
\label{sec:dataset}

We used traffic data for a specific range of IPv4 address blocks collected by the NICTER\footnote{\url{https://www.nicter.jp/en}} project\cite{4627315}, a packet monitoring project operated by the National Institute of Information and Communications Technology (NICT)\footnote{\url{https://www.nict.go.jp/en/index.html}} as the target traffic. The target traffic we selected covers a time period  from July 2020 to September 2020. There are several different types of transport protocols seen in the dataset; however, we only focus on TCP and UDP herein. The total number of packets in the dataset is shown in Table~\ref{tab:dataset-size}.

\begin{table}
    \centering
    \caption{The number of packets in the dataset}
    \label{tab:dataset-size}
    \begin{tabular}{c|r|r}
        \textbf{Month} &  \multicolumn{1}{c|}{\textbf{TCP}} & \multicolumn{1}{c}{\textbf{UDP}}\\
        \hline
        July 2020 & 1,277,502,885 & 82,871,491 \\
        August 2020 & 1,162,490,968 & 91,500,000 \\
        September 2020 & 1,394,414,668 & 103,852,293 
    \end{tabular}
\end{table}

\section{Data Cleansing}
\label{sec:data-cleansing}

Although our approach is straightforward, it does not provide reasonable results as far as we just simply use the raw port access statistics data. There are two important characteristics of the port access statistics data.

\begin{enumerate}
    \item Effects of accesses to popular ports
    \item Effects of a small number of constant daily accesses
\end{enumerate}

When using TF--IDF for natural language documents, we remove \textit{stop words} from the documents. Stop words are words frequently appear in the documents, but have no important meaning in the analysis context, such as ``a'' or ``the'' in English.

Internet traffic has a huge bias with regard to the usage of ports. For example, ports 80 and 443 are mostly used for web access and currently one of the major types of Internet traffic. Almost all of the Internet services are built on top of web technology. As a result, the access counts to these web related ports tend to be big. In this study, we do not focus on detection of anomaly behavior hidden in the large amount of traffic using such ports. Instead, we focus on other ports activities often used by IoT-based malware. We discuss how to decide the list of ports considered as stop words in Section~\ref{sec:filtering-major-ports}.

The second characteristic is important when we use the TF--IDF-based analysis.  As shown in Equation~\ref{eqn:tf-idf}, the value is calculated from the term frequency and inverse document frequency.  The latter roughly means how much special the word appearance is in the set of documents. If a word appears only in a specific document and does not appear in other documents, we can assume that the word has a special value in the document. If all the documents contains the same words, we cannot decide which word is a special word in a specific document. In documents written in natural languages, this does not happen usually, but not the case in network traffic.

Here we applied TF--IDF to the network port access history. A port number is a word, and a one-day access history is a document. In the network access log, especially when the scale of a network is large, most of the ports are observed daily. We face several scanning attacks almost every day. In some cases, such activities scan the entire port space. As a result, if we do not preprocess the raw access data, then we will see all the documents (one-day access history) have logs of all the ports. In this case, the IDF values for each port will have the same value, and the TF--IDF calculation loses any meaning.

To avoid this effect, we filtered out ports that have only a small number of access histories in one day from its daily access log. For example, assume that we have a threshold value of 1,000 accesses per day. If we see the number of accesses less than 1,000 to port X, then we consider that port X does not appear on that day. The threshold value depends on the amount of traffic and/or the usage of the network. This can be considered as a kind of a simple noise filtering. We will discuss how to determine the threshold value in Section~\ref{sec:filtering-minor-ports}.

\subsection{Selecting stop words}
\label{sec:filtering-major-ports}

Table~\ref{tab:port-ratio} shows the ratio of TCP port access statistics of the dataset we used.

\begin{table}
\centering
    \caption{TCP port access ratio in the dataset (top 10)}
    \label{tab:port-ratio}
    \begin{tabular}{c|c|c}
         \textbf{Port number} & \textbf{Ratio} & \textbf{Mainly used by}  \\
         \hline
         445  & 0.101 & Windows File Sharing\\ 
         23   & 0.071 & TELNET \\
         1433 & 0.026 & SQL Server \\
         22   & 0.022 & SSH \\
         21   & 0.019 & FTP \\
         80   & 0.015 & HTTP \\
         1723 & 0.010 & PPTP\\
         5555 & 0.009 & Android Debug Bridge\\
         81   & 0.008 & HTTP alternative \\
         8080 & 0.008 & HTTP alternative / Proxy
    \end{tabular}
\end{table}

The major port numbers observed in the dataset depend on how the source network, where the data are captured, is used. In this study, we used the darknet traffic of certain IPv4 address blocks. As there was no host behind the target IPv4 address space, no active users were present in the network. All the packets came from outside. This means that the port statistics trend reflects how the packet originators thought the target network services were built.

In the case of the dataset used herein, we excluded ports 445, 23, 22, 80 81, 8080, and 443. Not all the ports listed in Table~\ref{tab:port-ratio} were excluded; instead, the port not shown in the table (port 443) was excluded. The selection of excluded ports was not fully systematic. We decided on them based on our heuristic knowledge.

\subsection{Noise filtering}
\label{sec:filtering-minor-ports}

As discussed at the beginning of this section, we filter out ports from a one-day access history based on the predefined minimum number of accesses per day. In this section, we describe the systematic determination of the minimum number.

The core idea of TF--IDF is that the frequency of terms depends on the document content. As a result, when we apply TF--IDF to text documents, we will have different IDF values for terms when we have a wide variety of documents. In the context of text processing, we normally apply data cleansing operations. For example, the word ``a'' or ``the'' is removed before applying TF--IDF, as such words are quite common in all documents and it does not make much sense when calculating the IDF values of terms.

The goal of our analysis is to obtain ``interesting'' ports from access logs using the TF--IDF mechanism. To focus on such ports, we try to maximize the number of ports (which are equivalent to terms in a text processing context) having larger IDF values. To do this, we calculate the histogram of IDF values with different threshold values of minimum access numbers.

\begin{figure}
\begin{subfigure}[t]{0.45\columnwidth}
    \includegraphics[width=\columnwidth]{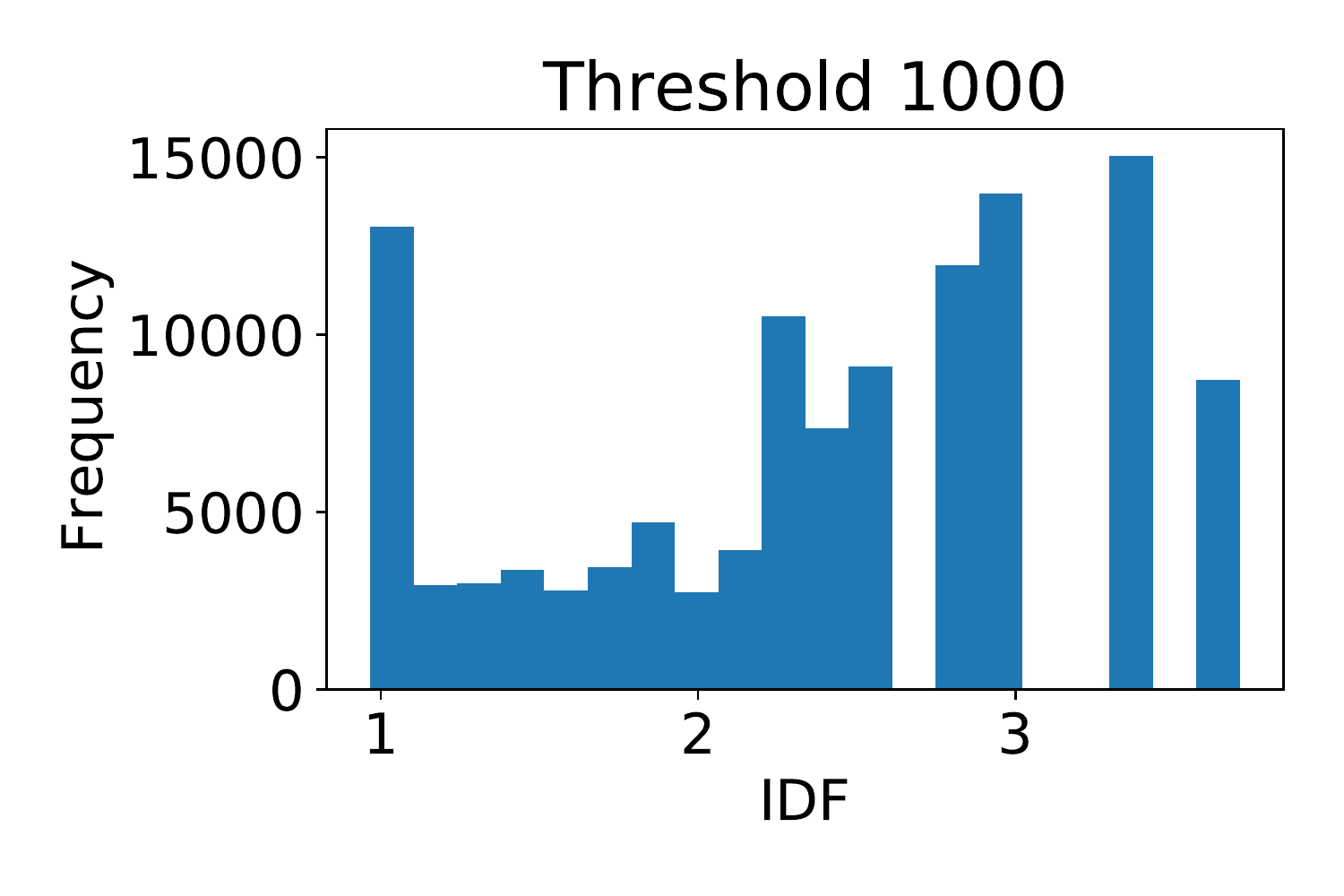}
    \caption{Threshold 1,000}
    \label{fig:t1000}
\end{subfigure}
\begin{subfigure}[t]{0.45\columnwidth}
    \includegraphics[width=\columnwidth]{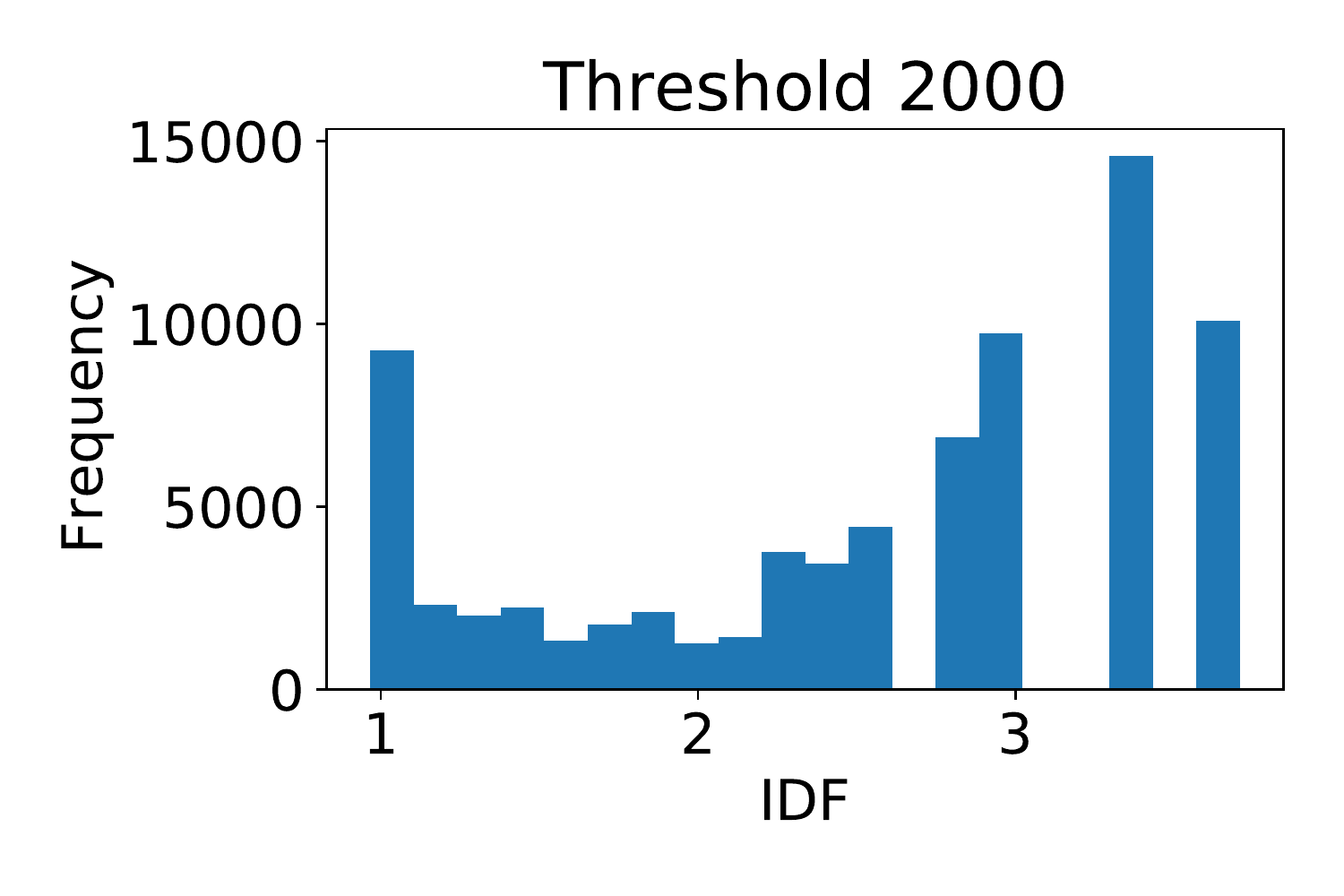}
    \caption{Threshold 2,000}
    \label{fig:t2000}
\end{subfigure}
\\
\begin{subfigure}[t]{0.45\columnwidth}
    \includegraphics[width=\columnwidth]{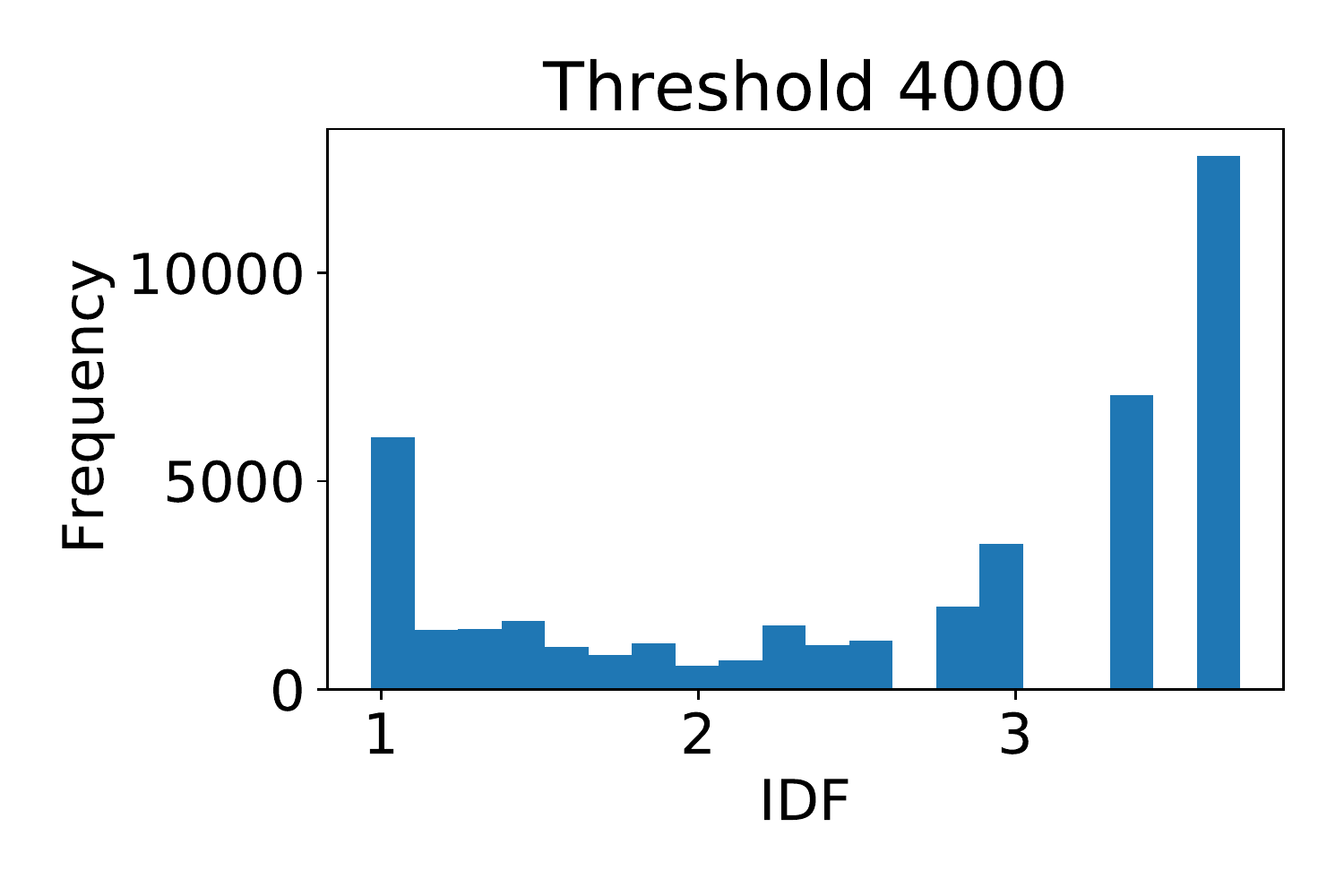}
    \caption{Threshold 4,000}
    \label{fig:t3000}
\end{subfigure}
\begin{subfigure}[t]{0.45\columnwidth}
    \includegraphics[width=\columnwidth]{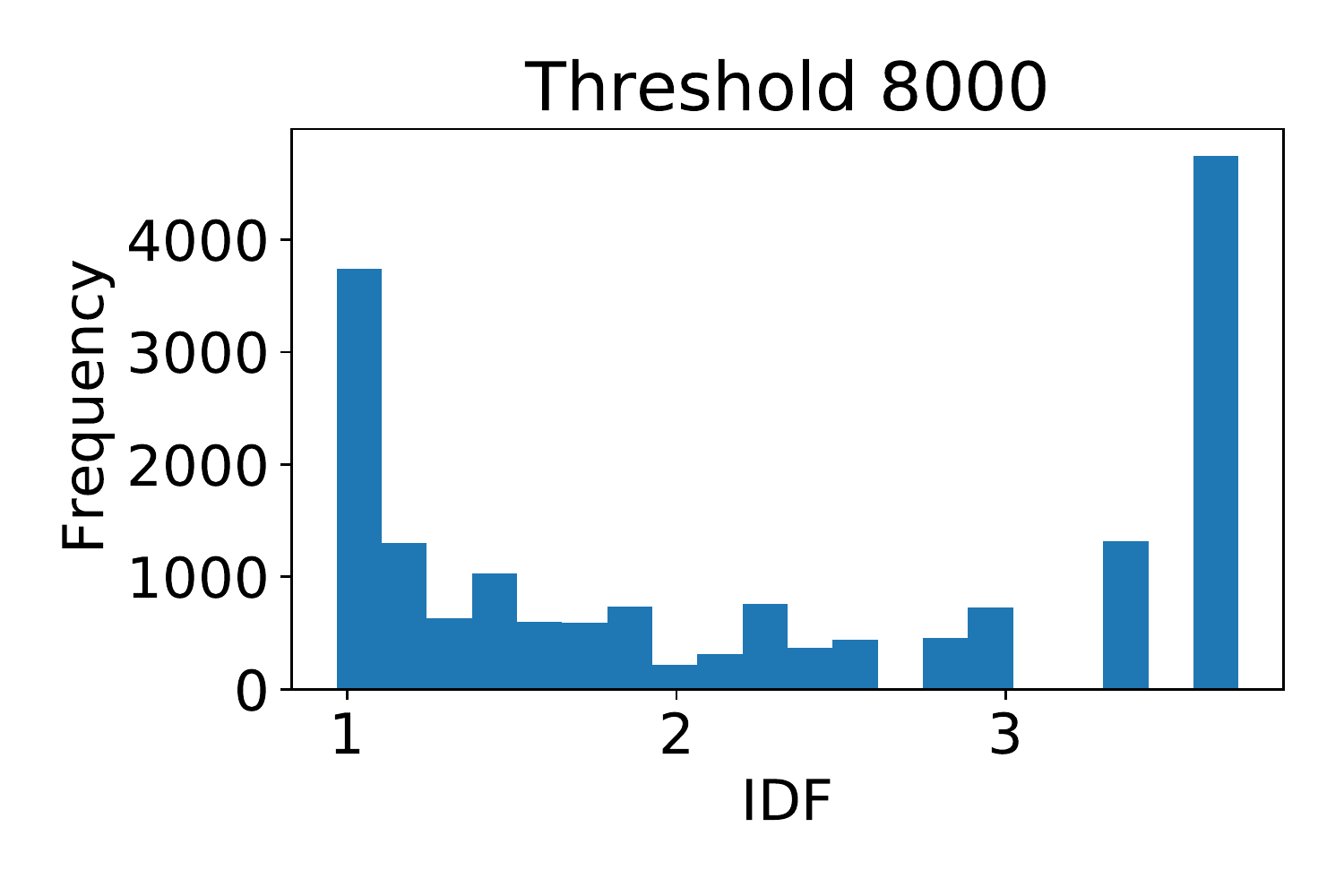}
    \caption{Threshold 8,000}
    \label{fig:t4000}
\end{subfigure}
\caption{Histograms of IDF values with different threshold values}
\label{fig:idf-histogram}
\end{figure}

Figure~\ref{fig:idf-histogram} shows the histograms of IDF value using different threshold values. The target data contains 30 days of access history, indicating 30 documents in the TF--IDF context. When a port is observed only in one day, the IDF value of the port will have the maximum value $log(N / (df(t) + 1)) + 1 = log(30 / (1 + 1)) + 1= 3.70...$ based on Equation~\ref{eqn:tf-idf}. If the threshold value is too small, we will see more minor ports in the access logs and we see fewer ports having higher IDF values. On the contrary, if we use a large threshold value, such minor ports are completely removed from the access logs, and we may miss the activities of minor ports. In Figure~\ref{fig:idf-histogram}, we used 4 different threshold values and we can confirm that the maximum number of ports having the largest IDF values can be observed when we use a threshold value of 4,000.

In the analysis parts described in the later sections, we use the above approach to determine the minimum threshold value. We start the value from 1,000 and calculate the histogram of the IDF values. Subsequently, the threshold value is doubled, and the histogram is recalculated. We stop the calculation when we see a decreasing trend in the maximum number of ports with the largest IDF value.

\section{TF--IDF results}
\label{sec:tf-idf-results}

In this section, we present some of the results of our TF--IDF-based port statistics analysis. As described in Section~\ref{sec:filtering-major-ports}, we exclude some of the port numbers treated as stop words in the dataset. The access logs of these ports were entirely removed from access history. In addition, as described in Section~\ref{sec:filtering-minor-ports}, ports having only a small number of accesses in one day are removed from the access log of that specific day.

The number of days (the number of documents in the TF--IDF context) to calculate the TF--IDF values was 30. When calculating the TF--IDF values of August 1, we used the access logs from July 3 to August 1. We calculated the TF--IDF values of each port from August 1 to September 30 and checked the access history of the ports that had high TF--IDF values.  The reason why we uses a sliding window-based approach is that anomaly behavior has short-term trend. We focus on activities of recent days (30 days in our study) to be able to find fresh events useful for a continuous daily operation.

\subsection{TCP port 9530}

\begin{figure*}[tbhp]
    \centering
    \includegraphics[width=0.9\textwidth]{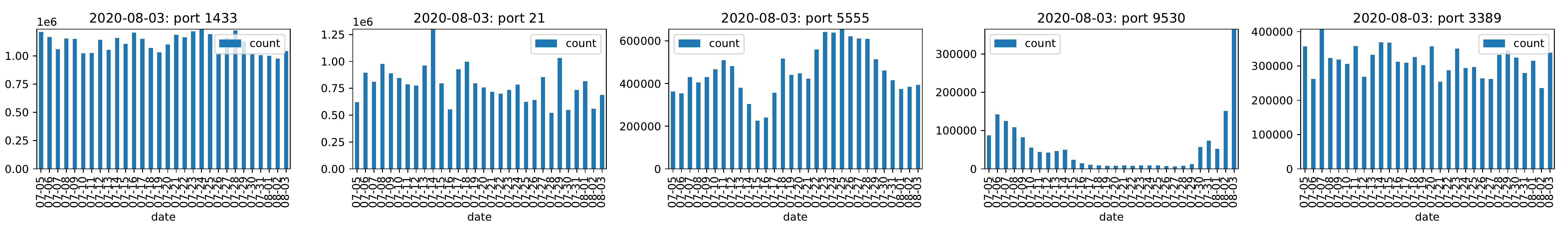}
    \caption{Past 30 days of access histories of ports listed as top-5 TF--IDF values on August 3, 2020}
    \label{fig:tfidf_tcp_2020-08-03}
\end{figure*}

Figure~\ref{fig:tfidf_tcp_2020-08-03} shows the past 30 days of access history of the top-5 TF--IDF values of TCP ports on August 3. We observe that the traffic on TCP port 9530 suddenly increased. The history shows that the activity started around July 30. 

\begin{figure*}[tbhp]
    \centering
    \includegraphics[width=0.9\textwidth]{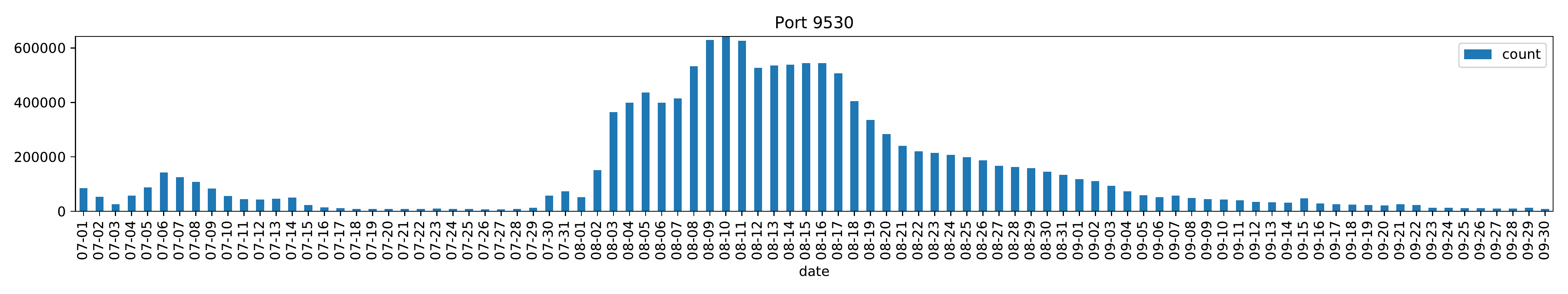}
    \caption{Access history of TCP port 9530 from July 1 to September 30, 2020}
    \label{fig:tcp_9530}
\end{figure*}

Figure~\ref{fig:tcp_9530} shows the access history of TCP port 9530 from July to September. The port is known as the port used by the subspecies of the Mirai bot, a well-known malware targeting mainly IoT devices\cite{Antonakakis2017}. The number of accesses to the port 9530 started increasing from the end of July, had a peak value in the middle of August, and gradually disappeared by the end of September. Such an activity related to the port 9530 is not a rare case. If we check the past access logs, we could find that traffic to port 9530 was observed several times. For example, one of the web reports mentions that a similar activity happened in February 2020\footnote{D4 Project Blog: \url{https://www.d4-project.org/2020/03/06/analyzer-d4-isn.html}}. Despite this fact, we argue that it is still worth finding such re-activated cases.

\begin{figure*}[tbhp]
    \centering
    \includegraphics[width=0.9\textwidth]{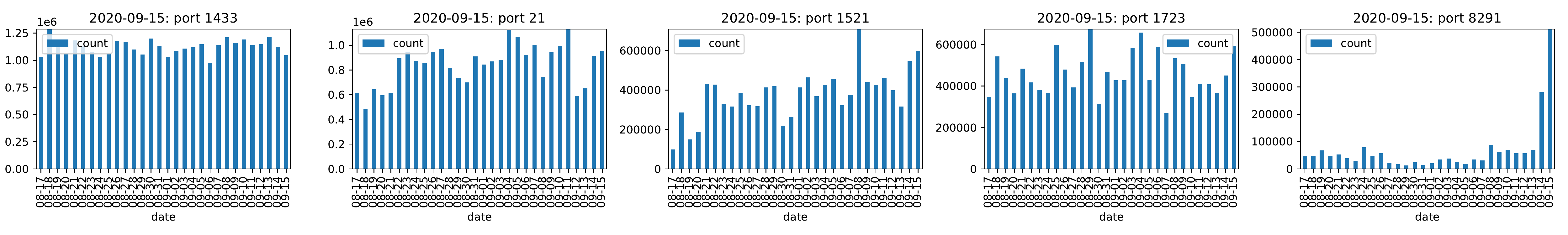}
    \caption{Past 30 days of access histories of ports listed as top-5 TF--IDF values on September 15, 2020}
    \label{fig:tfidf_tcp_2020-09-15}
\end{figure*}

The blog article also mentions a special characteristic of Mirai-based malware. The original Mirai software uses the same value for the destination IP address and the TCP initial sequence number when creating the first TCP packet. We also confirmed that the packets captured in our dataset had the same characteristics. This fact implies that we could conclude that the activity observed in August was related to the Mirai subspecies with high probability.

\subsection{TCP port 8291}
\label{sec:tcp-8291}

\begin{figure*}[tbhp]
    \centering
    \includegraphics[width=0.9\textwidth]{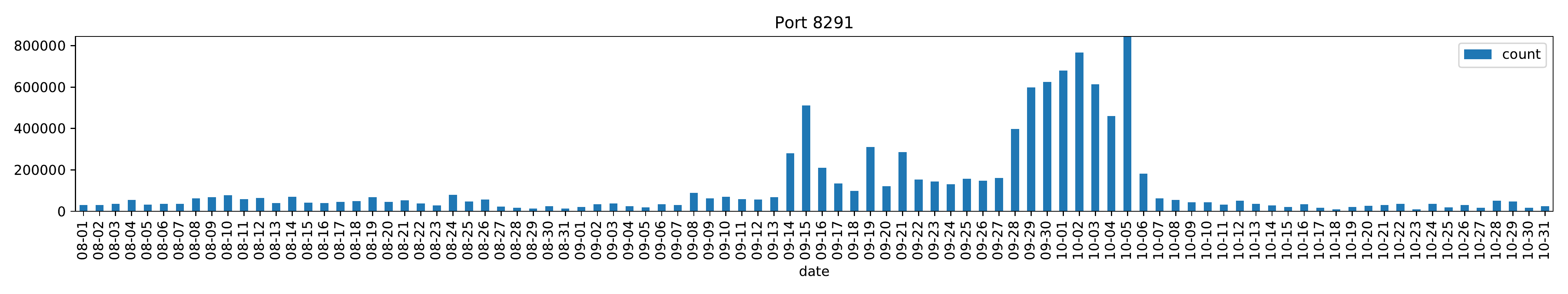}
    \caption{Access history of TCP port 8291 from September 1 to October 31, 2020}
    \label{fig:tcp_8291}
\end{figure*}

Figure~\ref{fig:tfidf_tcp_2020-09-15} is another example observed on September 15. On this day, the TCP port 8291 was listed as the port with the fifth-highest TF--IDF value. Figure~\ref{fig:tcp_8291} shows the activity of the port from August to October. We can observe the access to that port started increasing around September 8 and stopped around October 5. This port is known as the port used by the RouterOS produced by MikroTik. Some versions of the OS have vulnerability\cite{ceron2020characterising} reported as CVE-2019-3978\footnote{CVE-2019-3978: \url{https://cve.mitre.org/cgi-bin/cvename.cgi?name=CVE-2019-3978}}. In 2020, we observed the increase of access to port 8291 a few times, and this activity is considered one of them.

\subsection{UDP Waves}
\label{sec:udp-waves}

\begin{figure*}[tbhp]
    \centering
    \includegraphics[width=0.9\textwidth]{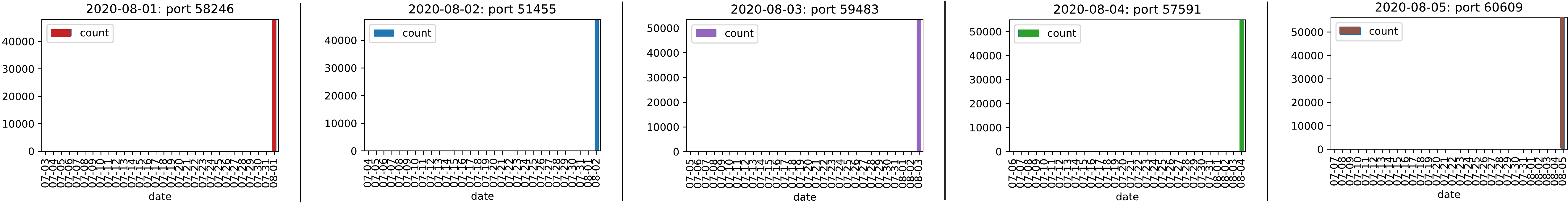}
    \caption{Access histories of the UDP ports with the largest TF--IDF values from August 1 to 5, 2020}
    \label{fig:tfidf_udp_202008}
\end{figure*}

\begin{figure*}[tbhp]
    \centering
    \includegraphics[width=0.9\textwidth]{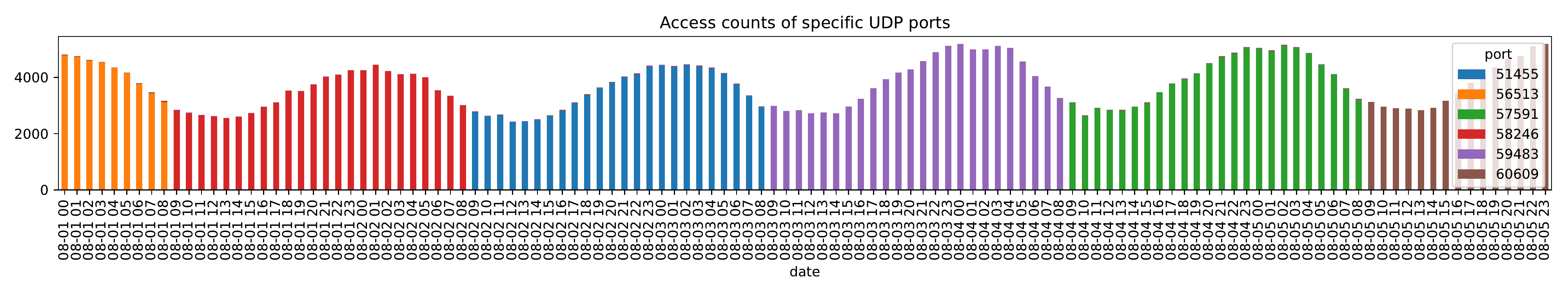}
    \caption{Combined access history of the UDP ports with the largest TF--IDF values from August 1 to 5, 2020}
    \label{fig:udp-wave}
\end{figure*}

The previous sections discussed the two TCP ports found by our proposed methods. We also analyzed UDP port access history. In the case of UDP, we did not find any specific trend of increased port numbers at a specific time frame; however, we observed that mysterious UDP accesses exist in the dataset. Figure~\ref{fig:tfidf_udp_202008} shows the ports which TF--IDF values were largest during August 1 and 5. As indicated by the TF--IDF value, these ports were observed mainly on that day. For example, the large amount of access to the UDP port 58246 was mostly observed on August 1 (and August 2, which is although not shown in the paper), the UDP port 51455 was observed mostly on August 2 (and August 3), and so on. Figure~\ref{fig:udp-wave} shows the hourly access history of the port numbers shown in Figure~\ref{fig:tfidf_udp_202008} from August 1 to 5. Access to a specific port starts daily at 09:00 in Japanese Standard Time (00:00 in UTC) and continues for 24 h. After finishing the access to the specific port, the other access to a different UDP port number starts and continues for 24 h. Interestingly, the port number was changed precisely at 09:00 JST (00:00 UTC) daily. This implies that the traffic was controlled by software. It is also interesting that the amount of hourly traffic is not constant and made a waveform. This implies that the activity was based on social activities such as day--night trends, working hours, or similar activities.

As the dataset is captured by the network operated as a darknet, we could not identify what type of communication was expected to these UDP accesses because we did not send any response to the incoming UDP packets. We tried investigating the contents of the UDP packets; however, the facts we could achieve were not many.

\begin{figure*}[tbhp]
\centering
\begin{subfigure}[t]{0.15\textwidth}
    \centering
    \includegraphics[width=\columnwidth]{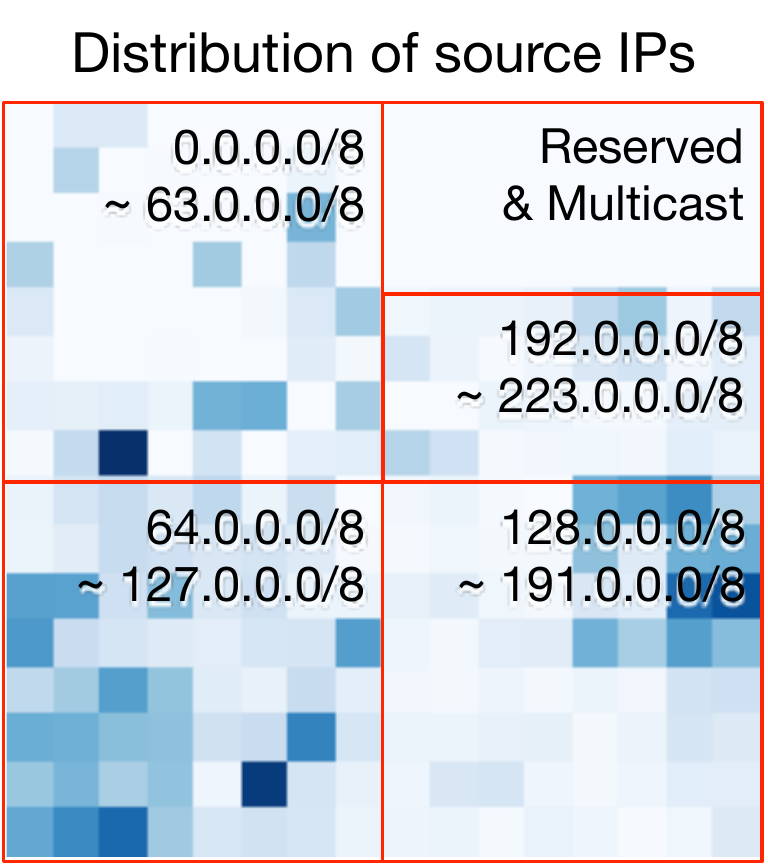}
    \caption{Heatmap of the source IP blocks in /8 using a Hilbert curve}
    \label{fig:udpwave_hilbert}
\end{subfigure}
\hspace{1em}
\begin{subfigure}[t]{0.23\textwidth}
    \centering
    \includegraphics[width=\columnwidth]{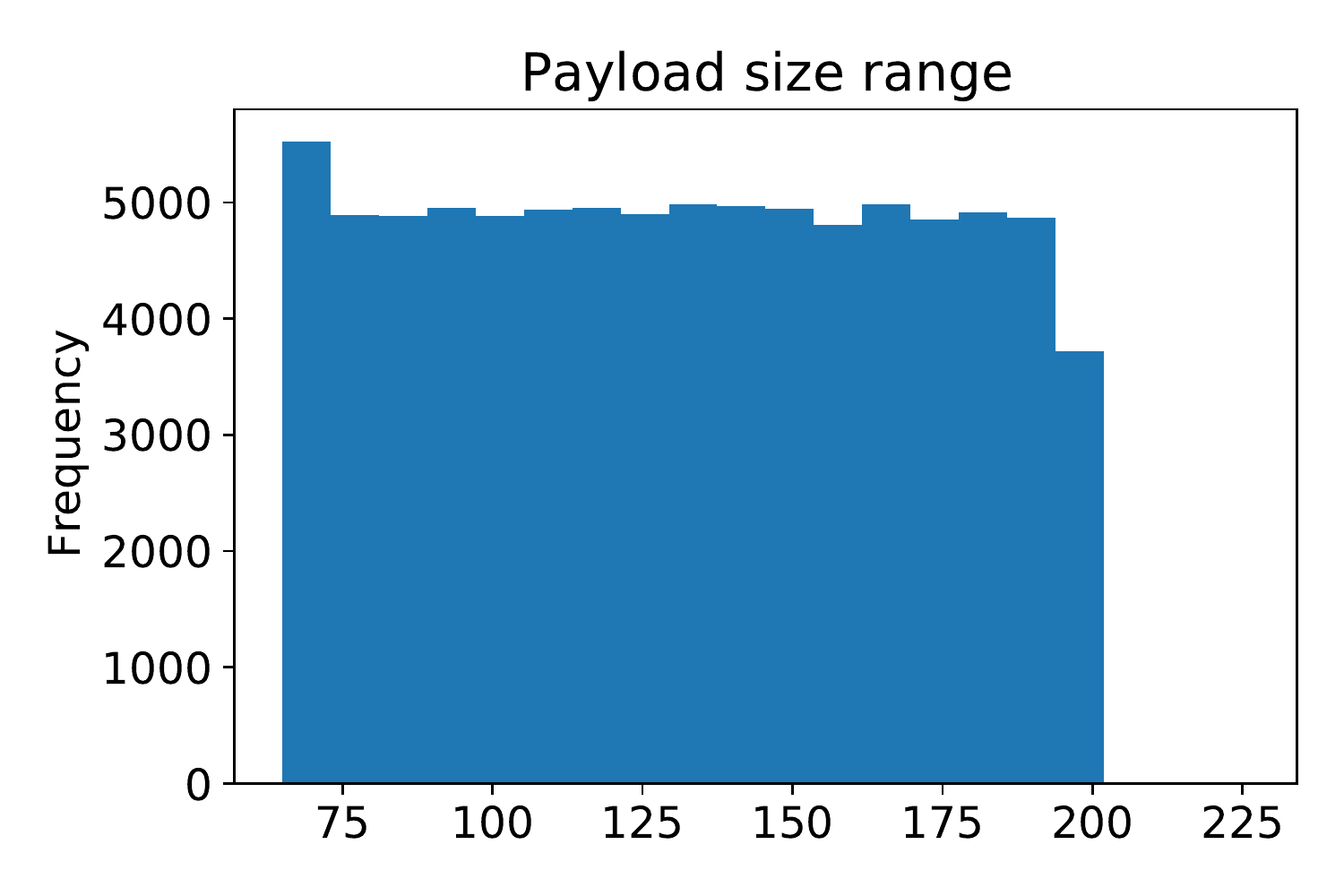}
    \caption{Distribution of the payload size}
    \label{fig:udpwave_plen}
\end{subfigure}
\hspace{1em}
\begin{subfigure}[t]{0.23\textwidth}
    \centering
    \includegraphics[width=\columnwidth]{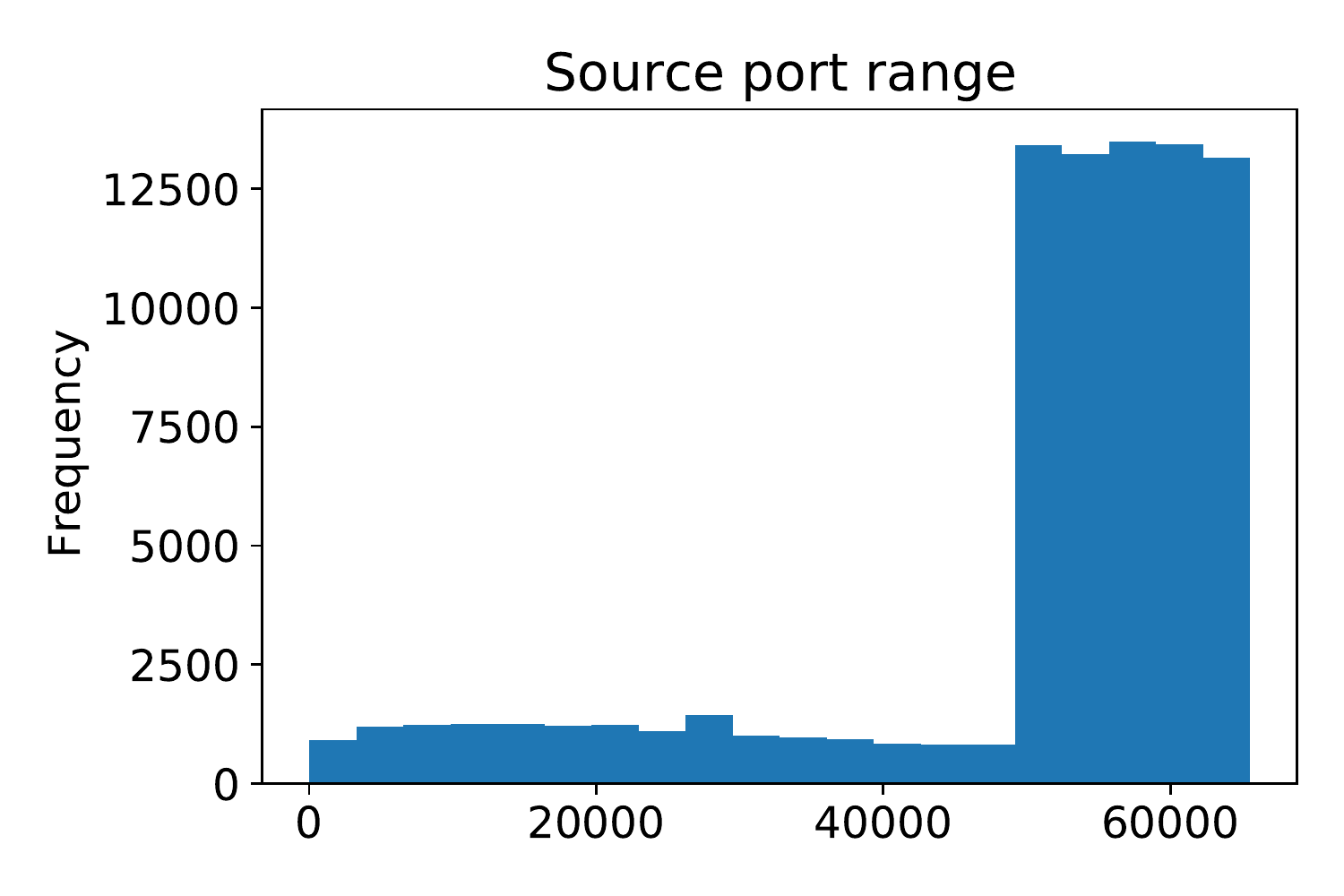}
    \caption{Distribution of the source port numbers}
    \label{fig:udpwave_port}
\end{subfigure}
\caption{The characteristics of the UDP wave traffic sent to the TCP port 58246 on August 1 and 2, 2020}
\end{figure*}

The source IP addresses of the incoming UDP packets were quite random, so we could guess that the source addresses were spoofed (perhaps some of them were correct source addresses, but we do not know how to verify the source IP addresses). Figure~\ref{fig:udpwave_hilbert} shows a heatmap of the number of source IP addresses sent to TCP port 58246 from August 1 and 2. The figure is drawn as a Hilbert curve, and each block indicates a /8 address block. The source IP addresses are distributed to the entire unicast IPv4 address space.

The size of the payload of the incoming UDP packets seemed to be random, ranging from 65 to 226 bytes. Figure~\ref{fig:udpwave_plen} shows the distribution of payload size. We briefly checked the contents of the payload; however, it seemed to be random bytes. There is a possibility that these data are some type of encoded control commands to the machines infected by malware; however, verifying the fact within our dataset is not possible.

The source port numbers are almost set to 50,000 or larger, as shown in Figure~\ref{fig:udpwave_port}. This range is normally used for ephemeral ports\cite{rfc6056}. In that sense, the packet generator of this UDP wave phenomenon respects the RFC standard recommendation, indicating that the generator probably uses normal packet generation APIs provided by the operating system.

\section{Discussion}

As shown in Section~\ref{sec:tf-idf-results}, we can observe some unusual traffic behaviors with our proposed method. However, we also noticed that there were activities not identified by this method. Before discussing such activities, we revisit the formula used to calculate the TF--IDF value in our proposal. The formula used to calculate the IDF value shown in Equation~\ref{eqn:tf-idf} is slightly different from that of the common formula used for the IDF calculation. The commonly used formula is shown in Equation~\ref{eqn:common-idf-formula}. The difference is the extra ``$+ 1$'' used in our proposal.

\begin{equation}
    \label{eqn:common-idf-formula}
    idf(t) = \log\frac{N}{df(t)+1}
\end{equation}

The formula we used herein is widely known as the smoothed IDF. We tried both the normal version and the smoothed version of the IDF formulae and observed that only the smoothed version provided us with meaningful results when used with our dataset. There were many activities that appeared only a day or two and did not appear in the remaining days. With the normal version, these activities will have higher TF--IDF values and hide ports discovered in this study.



With the above experience, we chose to use the smoothed version of the TF--IDF formula; however, it did not show the expected results in some cases. We are not sure how many cases there are, but we have confirmed that at least one case exists. In Section~\ref{sec:tcp-8291}, we demonstrate that the proposed method can detect the increasing trend of TCP port 8291. As explained in the section, we consider this activity to be related to the MikroTik RouterOS. We noticed that the port numbers reported in the related CVE was not only 8291 but also 8728 was often used with 8291. We checked the activity of TCP port 8728, which has not been reported with our mechanism.

\begin{figure*}[tbhp]
    \centering
    \includegraphics[width=0.9\textwidth]{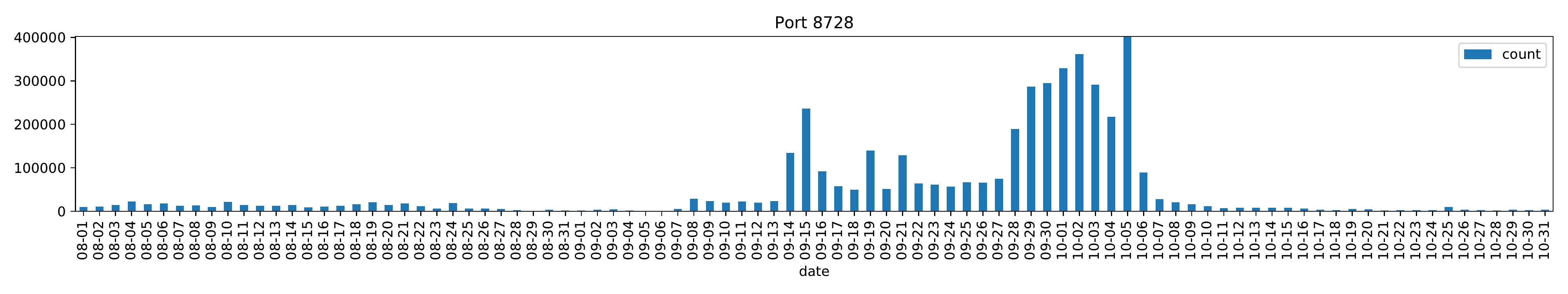}
    \caption{Undetected access history of TCP port 8728 from September 1 to October 31, 2020}
    \label{fig:tcp_8728}
\end{figure*}

Figure~\ref{fig:tcp_8728} shows the access history of TCP port 8728. We can confirm that accesses to ports 8291 and 8728 were synchronized, although the number of packets was different. The traffic to port 8728 was almost half of the volume compared to that of port 8291. We could not identify this port because the volume of the traffic was too small to raise the TF value of that port. We tried another smoothing approach that applies the logarithm when calculating the TF value; however, this did not improve the results.

\section{Related works}

The simplest way to obtain anomaly behavior based on time-series data is a threshold-based approach. This works well when the range of counts is predictable. On the Internet, the number of incoming packets is not under our control, and setting a proper threshold is impossible. 

The k-nearest neighbor and ARIMA are popular methods for detecting anomalies without knowing the normal range of the target time-series data\cite{8328485,SU20113492,4631947,5437603,7814437,8530608}. These approaches work well when a series of data continuously takes similar values. The port count data is, however, a spiky data series, and the range of the counts is sometimes unstable. We can ease these spikes by using a moving average or summing up the one-day data to make the data almost stable time-series data; however, we eventually see many anomaly results with these approaches because they do not consider how important each port count is compared to other port counts. The TF--IDF-based approach shows the relative importance among the port counts based on the frequency of port appearance.

\section{Conclusion}

We applied the TF--IDF algorithm, which is used in a natural language processing context, for observing the port access trend of Internet traffic. Using the real-life dataset, we identified two anomaly TCP port access behaviors and one periodical UDP access behavior from the dataset. As the proposed mechanism gives us a weighted order based on the popularity of port access frequency, we can focus on more important activities than using simple change detection mechanisms for each port activity. It is true that there is no single best mechanism to identify malicious activities; however, we hope that our approach can be a collaborative algorithm to obtain anomaly behaviors in conjunction with other approaches.

\section*{Acknowledgment}
We thank the NICTER project and the WIDE project\footnote{\url{https://www.wide.ad.jp/index_e.html}} for supporting our work.

\bibliographystyle{unsrt}
\bibliography{bibliography.bib}

\begin{thebibliography}{10}

\bibitem{4627315}
D.~{Inoue}, M.~{Eto}, K.~{Yoshioka}, S.~{Baba}, K.~{Suzuki}, J.~{Nakazato},
  K.~{Ohtaka}, and K.~{Nakao}.
\newblock nicter: An incident analysis system toward binding network monitoring
  with malware analysis.
\newblock In {\em 2008 WOMBAT Workshop on Information Security Threats Data
  Collection and Sharing}, pages 58--66, 2008.

\bibitem{Antonakakis2017}
Manos Antonakakis, Tim April, Michael Bailey, Matthew Bernhard, Ann Arbor, Elie
  Bursztein, Jaime Cochran, Zakir Durumeric, J~Alex Halderman, Ann Arbor, Luca
  Invernizzi, Michalis Kallitsis, Merit Network, Zane Ma, Joshua Mason, Damian
  Menscher, Chad Seaman, Nick Sullivan, Kurt Thomas, Yi~Zhou, Manos
  Antonakakis, Tim April, Michael Bailey, Matthew Bernhard, Elie Bursztein,
  Jaime Cochran, Zakir Durumeric, J~Alex Halderman, Luca Invernizzi, Michalis
  Kallitsis, Deepak Kumar, Chaz Lever, Zane Ma, Joshua Mason, Damian Menscher,
  Chad Seaman, Nick Sullivan, Kurt Thomas, and Yi~Zhou.
\newblock {Understanding the Mirai Botnet}.
\newblock In {\em USENIX Security}, pages 1093--1110, 2017.

\bibitem{ceron2020characterising}
Joao~M. Ceron, Christian Scholten, Aiko Pras, Elmer Lastdrager, and Jair
  Santanna.
\newblock Characterising attacks targeting low-cost routers: a mikrotik case
  study (extended), 2020.

\bibitem{rfc6056}
Michael Larsen and Fernando Gont.
\newblock {Recommendations for Transport-Protocol Port Randomization}.
\newblock RFC 6056, January 2011.

\bibitem{8328485}
Shengchu Zhao, Wei Li, Tanveer Zia, and Albert~Y. Zomaya.
\newblock A dimension reduction model and classifier for anomaly-based
  intrusion detection in internet of things.
\newblock In {\em 2017 IEEE 15th Intl Conf on Dependable, Autonomic and Secure
  Computing, 15th Intl Conf on Pervasive Intelligence and Computing, 3rd Intl
  Conf on Big Data Intelligence and Computing and Cyber Science and Technology
  Congress(DASC/PiCom/DataCom/CyberSciTech)}, pages 836--843, 2017.

\bibitem{SU20113492}
Ming-Yang Su.
\newblock Real-time anomaly detection systems for denial-of-service attacks by
  weighted k-nearest-neighbor classifiers.
\newblock {\em Expert Systems with Applications}, 38(4):3492--3498, 2011.

\bibitem{4631947}
H.~Zare~Moayedi and M.A. Masnadi-Shirazi.
\newblock Arima model for network traffic prediction and anomaly detection.
\newblock In {\em 2008 International Symposium on Information Technology},
  volume~4, pages 1--6, 2008.

\bibitem{5437603}
Asrul~H. Yaacob, Ian~K.T. Tan, Su~Fong Chien, and Hon~Khi Tan.
\newblock Arima based network anomaly detection.
\newblock In {\em 2010 Second International Conference on Communication
  Software and Networks}, pages 205--209, 2010.

\bibitem{7814437}
Eduardo H.~M. Pena, Marcos V.~O. de~Assis, and Mario~Lemes Proença.
\newblock Anomaly detection using forecasting methods arima and hwds.
\newblock In {\em 2013 32nd International Conference of the Chilean Computer
  Science Society (SCCC)}, pages 63--66, 2013.

\bibitem{8530608}
Rishabh Madan and Partha~Sarathi Mangipudi.
\newblock Predicting computer network traffic: A time series forecasting
  approach using dwt, arima and rnn.
\newblock In {\em 2018 Eleventh International Conference on Contemporary
  Computing (IC3)}, pages 1--5, 2018.

\end{thebibliography}

\end{document}